\documentclass[aps,pre,superscriptaddress,
               twocolumn,balancelastpage,
               ]{revtex4-1}

\usepackage[colorlinks,bookmarks=false,citecolor=blue,linkcolor=blue,urlcolor=blue]{hyperref}
\usepackage[all]{hypcap}   

\usepackage{amsmath,amssymb}
\usepackage{graphicx}

\usepackage{verbatim}
\usepackage{color}

\usepackage{placeins}  
\usepackage{flafter}     

\usepackage{braket}
\usepackage{dsfont}

\renewcommand{\mod}{\text{mod }}

\newcommand{\ue}{\text{e}}
\newcommand{\ui}{\text{i}}
\newcommand{\ud}{\text{d}}
\newcommand{\U}{\mathcal{U}}
\newcommand{\T}{\mathcal{T}}
\newcommand{\M}{\mathcal{M}}
\newcommand{\Proj}{\mathcal{P}}
\newcommand{\Hil}{\mathcal{H}}
\newcommand{\tr}{\text{tr}}

\newcommand{\Uloc}[8]{\tiny $\left( \begin{smallmatrix}
        #1  \ui #2  & #3 \ui #4   \\
         #5 \ui #6  & #7 \ui  #8
    \end{smallmatrix}
\right)$ \normalsize
}

\makeatletter
\let\Hy@backout\@gobble
\makeatother

\begin{document}

\title{Eigenstate thermalization in dual-unitary quantum circuits:  \\
    Asymptotics of spectral functions
}

\author{Felix Fritzsch}
\affiliation{Physics Department, Faculty of Mathematics and Physics,
    University of Ljubljana, Ljubljana, Slovenia}

\author{Toma\v{z} Prosen}
\affiliation{Physics Department, Faculty of Mathematics and Physics,
    University of Ljubljana, Ljubljana, Slovenia}

\date{\today}

\begin{abstract}
    The eigenstate thermalization hypothesis provides to date
    the most successful description of
    thermalization in isolated quantum systems by conjecturing statistical properties of
    matrix elements of typical operators in the (quasi-)energy eigenbasis.
    Here we study the distribution of matrix elements for a class of operators in dual-unitary quantum circuits in dependence of the frequency associated with the corresponding eigenstates.
    We provide an exact asymptotic expression for the spectral function, i.e., the second moment of this frequency resolved distribution.
    The latter is obtained from the decay of dynamical
    correlations between local operators which can be computed exactly from the elementary building blocks of the dual-unitary circuits.
    Comparing the asymptotic expression with results obtained by exact diagonalization we find excellent agreement.
    Small fluctuations at finite system size are explicitly related to dynamical correlations at intermediate times and the deviations from their asymptotical dynamics.
    Moreover, we confirm the expected Gaussian distribution of the matrix elements by computing higher moments numerically.
\end{abstract}

\maketitle

\section{Introduction}

Statistical mechanics provides an accurate description of large quantum systems
in thermal equilibrium.
How isolated quantum systems approach thermal equilibrium, however, 
has posed major challenges for our understanding of 
quantum dynamics and statistical physics.
In particular, unitarity of time evolution seems to contradict the notion of 
thermalization from an out-of-equilibrium initial state.
This issue was addressed early on by
von Neumann \cite{Neu1929}, who proposed to study typical or macroscopic
physical observables only.
Combining this point of view with ideas from random matrix 
theory lead to the celebrated eigenstate thermalization hypothesis (ETH) which 
conjectures universal statistical properties of matrix elements of large classes of observables in
dependence on the respective energies \cite{Deu1991, Sre1994, Sre1999}.
Since then both the validity of ETH and its breakdown as well as the respective 
implications on thermalization have been investigated
in a multitude of studies \cite{Rei2015, AleKafPolRig2016, Deu2018, MorIkeKamUed2018}.
They are driven by recent experimental realizations of nonequilibrium many-body quantum dynamics 
\cite{GreManHaeBlo2002, TroCheFleMcCSchEisBlo2012, 
    KinWenWei2006, GriKuhLanKitRauSchMazSmiDemSch2012, 
    LanErnGeiRauSchKuhRohMazGasSch2015, SchHodBorLueFisVosAltSchBlo2015, 
    ChoHilZeiSchRubYefKheHusBloGro2016,KauTaiLukRisSchPreGre2016, 
    NeiRouFanCheKolCheMegBarCam_et_al2016, 
    TanKaoLiSeoMalRigGopLev2018,BluOmrLevKeeSem_etal2021}.

As we will focus on Floquet systems, which evolve in discrete time, 
let us formulate ETH explicitly in this context.
There ETH conjectures properties of matrix elements $A_{mn}=\bra{m}A\ket{n}$ 
of an observable $A$ in the 
quasi-energy eigenbasis $\{\ket{n}: \U\ket{n} = \ue^{\ui \varphi_n} \ket{n}\}$,
where $\U$ is the time evolution operator over one period -- the so-called Floquet operator.
For many-body systems with a clear spatial locality structure, 
i.e., being defined on a regular lattice with local interactions 
one typically considers $A$ from a class of a local observables 
or extensive sums thereof.
The ETH ansatz in this setting reads
\begin{align}
A_{mn} = \langle A \rangle\delta_{mn} + D^{-1/2}f(\omega_{mn})R_{mn}.
\label{eq:ETH}
\end{align}
Here, $\langle \cdot  \rangle$ denotes the thermal average, which in the case of quantum 
circuits or Floquet systems is taken with respect to the only generally invariant state 
-- the infinite temperature state: $\langle A\rangle = \frac{1}{D} \tr A$, $D=\tr \mathds{1}$.
In particular, there is no explicit dependence on the eigenphases, i.e., the quasi-energies
$\varphi_n$. 
Without loss of generality the thermal average might be set to zero by subtracting 
$\langle A\rangle  \mathds{1}$ from the operator $A$.
Consequently, the nontrivial information is encoded in the second term in which $D$ 
represents the dimension of the underlying Hilbert space,
$R_{mn}$ denote random variables with zero mean and unit variance, and $f(\omega)$ is
the spectral function, also called the structure function, of the observable $A$.
The latter is a smooth function of the eigenphase 
differences $\omega_{mn}=\varphi_n - \varphi_m$. 
Thus the central questions are:
(i) what is the underlying distribution of the random variables $R_{mn}$ or, 
equivalently,  of matrix elements $A_{mn}$, and (ii), what is the functional form of the 
spectral function $f(\omega)$?

These questions are interrelated and have been investigated in many studies 
with most of them focusing on the Hamiltonian (continuous-time)
case rather than Floquet systems.
For instance, the distribution of both diagonal and 
off-diagonal matrix elements has been confirmed to 
be generically well described by a Gaussian distribution in the quantum ergodic (or quantum chaotic) regime,
whose variance decreases exponentially when increasing the system size
\cite{BirKolLae2010, BreLeBGooRig2020, LeBRig2020, BeuMoeHaq2015,
SchJanHeiVid2020:p, JanStoVidHei2019, Noh2021:p, 
RicDymSteGem2020, BreLeBGooRig2020, LeBMalVidRig2019, BreGooRig2020}.
For these distributions the ratio of the variance of diagonal and off-diagonal matrix 
elements agrees with random matrix predictions 
\cite{MonRig2017, BreLeBGooRig2020, LeBRig2020, BreGooRig2020}.
Deviations from the Gaussian nature of the distribution have been found when the system 
under consideration approaches the integrable or localized regime 
\cite{Rig2009, RoyLevLui2018, BreLeBGooRig2020, LeBMalVidRig2019, BreGooRig2020} 
as well as for specific nonlocal operators \cite{KhaHaqMcC2019}.
Moreover, the differences between diagonal matrix elements of
neighboring eigenstates as well as
the difference between diagonal matrix elements and the corresponding 
microcanonical expectation value in general vanish with increasing
system size \cite{RigDunOls2008, SanRig2010, RigSan2010, KimIkeHus2014, MonFraSreRig2016, BeuMoeHaq2014, YosIyoSag2018, 
    SteKhoNieGogGem2014, MieVid2020, SugHamUed2020:p}.

More details on the distribution of matrix elements have
been obtained by studying the 
spectral function $f(\omega)$ as it encodes the variance of the distribution of matrix 
elements \cite{NatPor2018b, JanStoVidHei2019, RicDymSteGem2020,
    MonRig2017, LeBMalVidRig2019, LeBRig2020} and as it sets
the energy scale above which nontrivial correlations become relevant
\cite{Dym2018:p}.
Generalizations of the spectral function additionally yield higher order correlations 
of matrix elements \cite{ChaDeCha2019,FoiKur2019,BrePapMitGooSil2021:p}.
While the small frequency behavior $\omega \to 0$ of the
spectral function and the corresponding 
statistics of diagonal matrix elements $A_{nn}$ encodes equilibrium properties
at large times $t \to \infty$, finite frequencies $|\omega|>0$ and the statistics of 
off-diagonal elements determine fluctuations in equilibrium as well
as the dynamics of relaxation to equilibrium.
Consequently, the spectral functions governs the decay of dynamical
correlations via fluctuation-dissipation relations and linear response theory 
\cite{LuiBar2016, RicGemSte2019, FoiCugGam2012, KhaPupSreRig2014, NatPor2019, 
    NohSagYeo2020} yielding, e.g., heating rates in driven systems 
\cite{MalRig2019} and sensitive probes to quantum chaos
\cite{PanClaCamPolSel2020, SelPol2020:p}.

The aforementioned findings impressively demonstrate the validity of the ETH
given by Eq.~\eqref{eq:ETH} 
in generic systems accessible by numerical methods.
Exact results, however, are rare due to the complex and analytically
intractable dynamics of typical many-body systems.
This has triggered the search for exactly solvable chaotic models, from which
unitary quantum circuits emerged as a promising candidate.
One may hope to derive exact statements on the validity of the
ETH as these models allow for analytic results on various other fundamental 
properties of generic many-body quantum systems.
The latter include,
e.g., the ballistic spreading of local operators \cite{KheVisHus2018, 
    RakPolvon2018,NahVijHaa2018,ChaLucCha2019,ChaLucCha2018b,
    BerKosPro2019b,vonRakPolSon2018,DiaHaqRibMcC2021:p}
and the growth of entanglement
\cite{NahRuhVijHaa2017,BerKosPro2019,ChaLucCha2018b,SkiRuhNah2019,
    RakPolKey2019,GopLam2019,PirBerCirPro2020,BerKosPro2020a,BerKosPro2020b, KloBerPir2021:p, BerPir2020}
as well as random matrix predictions of spectral correlations
\cite{ChaLucCha2018a, ChaLucCha2018b,KosLjuPro2018,
     BerKosPro2018,FriChaDeCha2019,FlaBerPro2020,KosBerPro2020:p,BerKosPro2021:p}
and random state entanglement \cite{HamSanZan2012b,HamSanZan2012a}.
A particularly fruitful approach assumes an additional duality symmetry between space and time 
first observed \cite{AkiWalGutGuh2016} in a quantum chaotic Ising chain in a pulsed magnetic field \cite{Pro2002}. 
This lead to the notion of dual-unitary quantum circuits \cite{BerKosPro2019b} in which
both propagation in time as well as in space is unitary.
Dual-unitarity allows for analytical computations of 
dynamical correlation functions and 
spectral correlations as well as the spreading of local operators  
\cite{BerKosPro2018,BerKosPro2019,BerKosPro2019b,PirBerCirPro2020,
    BerKosPro2020a,BerKosPro2020b,ClaLam2020:p, BerKosPro2021:p,
    KosBerPro2021, FlaBerPro2020, ReiBer2021:p} and provides a manifestation of maximally entangling local evolutions
 \cite{Arul1,Arul2}.
Morover, the dual space evolution provides deep insights into non-equilibrium properties 
of general unitary as well as non-unitary
Floquet quantum circuits \cite{Vedika1,Vedika2,Dima1,Hamazaki, LuGro2021:p}.
    
Here we aim for analytical answers to the questions on the 
statistics of matrix elements and the form of the spectral function
within the setting of dual-unitary quantum circuits.
More precisely, we derive the
exact asymptotic form of the spectral function for large system sizes for a class of 
observables which comprise of sums of local operators.
To this end, we analytically compute their autocorrelation functions
for times proportional to system size
which are fully determined by a completely positive trace preserving (CPTP) map 
acting on local operators \cite{BerKosPro2019b}.
The latter is constructed using only the elementary building blocks, i.e., the local gates
from which the quantum circuit is built.
Its spectral properties yield the decay of correlations and asymptotically
govern the spectral function.
In particular, we find a Lorentzian shape of the 
spectral function in the presence
of isolated slowly decaying modes, while the spectral function becomes almost flat
when correlations decay fast. 

Using exact diagonalization of dual-unitary qubit circuits we compare the asymptotical 
result with numerically computed spectral functions and find excellent agreement for 
generic cases.
Nevertheless, we observe finite size deviations which we relate to the dynamics of 
autocorrelations on intermediate time scales and which are exponentially suppressed
when increasing the system size.
Going beyond what we can access analytically we 
additionally extend our numerical studies to higher moments of the 
frequency resolved empirical distribution
of matrix elements.
We confirm the latter to be well described by a Gaussian distribution.
However, disregarding the frequency dependence we find deviations from a Gaussian 
distribution.
More precisely, we observe exponential tails when the corresponding circuit exhibits 
slowly decaying modes. 

The remainder of this paper is organized as follows.
In Sec.~\ref{sec:circuits} we introduce the quantum circuits studied in our work.
Moreover, we review the concept of dual-unitarity and its implications on dynamical
correlations of local operators.
Subsequently, in Sec.~\ref{sec:spectral_function}, we provide details on the computation
of the spectral functions and its relation to the dynamics of autocorrelation functions.
We also introduce a class of operators for which the latter can be computed exactly.
As our main result we derive the asymptotic form of the spectral function
for these operators.
This is compared with numerical simulations in Sec.~\ref{sec:numerics}.
We additionally study the distribution of matrix elements there.
Eventually, we conclude and summarize our results in Sec.~\ref{sec:conclusions}.

\section{dual-unitary Quantum Circuits \label{sec:circuits}}

\begin{figure}[t!]
    \includegraphics{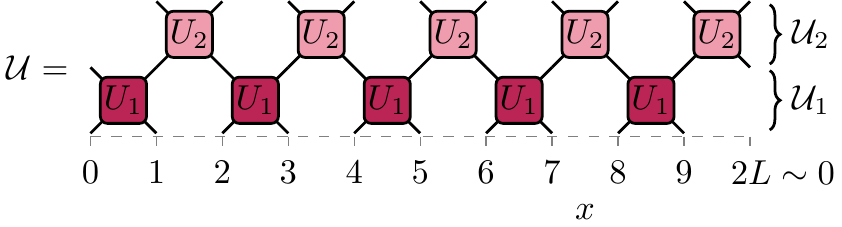}
    \caption{Diagrammatic representation of 
        $\U = \U_2\U_1$ with $\U_1$ and $\U_2$ given by Eq.~\eqref{eq:U_12} for $2L=10$.
        \label{fig:U_diagramm} }
\end{figure}

In the following we briefly review dual-unitary quantum circuits and some or their
fundamental dynamical properties.
In particular we focus on the dynamics of correlation functions of local operators.

\subsection{Circuit design}

We consider a one-dimensional lattice of qudits subject to a discrete time evolution
obtained from a brick-wall circuit design as illustrated in Fig.~\ref{fig:U_diagramm}.
Each qudit is described by a local Hilbertspace $\Hil_x = \mathbb{C}^q$ of dimension
$q$ indexed by the sites $x \in \{0, \ldots, 2L-1\}$.
The latter are arranged on a one-dimensional chain of even length $2L$ with
periodic boundary conditions.
The system's total Hilbert space is 
$\Hil = \bigotimes_{x=0}^{2L-1}\Hil_x$ and has dimension $\text{dim}\Hil = D = q^{2L}$.
The chain undergoes 
discrete time evolution governed by the unitary (Floquet) operator $\U$.
More precisely, each time step is composed of two half-time steps
such that $\U = \U_2\U_1$ where the half steps are given by
 \begin{align}
\U_1 = \bigotimes_{x=0}^{L-1}U_1 \quad \text{and} \quad
\U_2 = \T\left(\bigotimes_{x=0}^{L-1}U_2\right)\T^{-1}, \label{eq:U_12}
\end{align}
see Fig.~\ref{fig:U_diagramm} for a diagrammatic representation of the resulting time evolution operator $\U$.
Here, $U_1,U_2 \in \text{U}(q^2)$ describe the interaction of qudits on even sites with
their right and left nearest neighbor, respectively.
In order to avoid undesired symmetries we choose $U_1 \neq U_2$.
Moreover, $\T$ denotes a periodic shift
on the lattice defined in the canonical basis by $\T \left(\ket{i_0}\otimes\ket{i_{1}}\otimes\ldots\otimes\ket{i_{2L-1}}\right)=
\ket{i_{2L-1}}\otimes\ket{i_{0}}\otimes\ldots\otimes\ket{i_{2L-2}}$, where $\{\ket{i},i=0,\ldots,q-1\}$ is an orthonormal basis of $\Hil_x$.
By construction, $\left[\T^2, \U\right]=0$ and consequently 
both $\U$ and $\T^2$ can be diagonalized simultaneously yielding common eigenstates
$\ket{n}$ and
their respective eigenphases, i.e., 
\begin{align}
\U\ket{n} = \ue^{\ui \varphi_n} \ket{n}  \quad \text{and} \quad
\T^2\ket{n} = \ue^{\ui \frac{2\pi k_n}{L}} \ket{n},
\end{align}
with (quasi-) momenta $k_n \in \{0,\ldots,L-1\}$ as
$\T^{2L} = \mathds{1}_{\Hil}$.

Despite their simple structure quantum circuits similar to those described above
are tractable by exact analytical methods only in certain limiting cases.
Here, we require that  the local gates $U_1$ and $U_2$ satisfy a property dubbed {\em dual-unitarity}
in order to allow for analytic calculations.
To this end, for a given local two-qudit gate $U$ we assign the dual gate $\tilde{U}$ 
defined by a reshuffling of matrix entries as \cite{BerKosPro2019b}
\begin{align}
\label{eq:dual_gate}
\bra{i_1}\otimes\bra{j_1}\tilde{U}\ket{i_2}\otimes\ket{j_2} := \bra{j_2}\otimes\bra{j_1}U\ket{i_2}\otimes\ket{i_1}.
\end{align}
The original gate $U$ is then called dual-unitary, if its dual $\tilde{U}$ is unitary as well, i.e.
$\tilde{U} \in U(q^2)$.
Note that for a generic local gate its dual is in general not unitary.
In what follows we assume both local gates $U_1$ and $U_2$ to be dual-unitary.
While the local gates $U_i$ are the elementary building blocks of the time evolution operator
$\U$ their duals can be thought of as the elementary building blocks of a spatial transfer
matrix describing propagation in space rather then time.
Dual-unitarity implies that propagation in space is unitary as well and thus gives rise to a structural
symmetry between space and time.

Whereas the complete characterization or classification of the set of dual-unitary gates is an open problem for general $q$, it has been solved for $q=2$ in Ref.~\cite{BerKosPro2019b}.
There it has been shown that the dual-unitarity condition fixes only $2$ out of $16$ parameters specifying an arbitrary unitary two-qubit gate $U\in U(4)$.
Hence dual-unitarity is not as restricting as one might think naively.

\subsection{Dynamical correlations of local operators}

The symmetry between space and time due to dual-unitarity imposes
strong restrictions on the form of dynamical correlations [see 
definition~\eqref{eq:cfdef} below] of local operators.
Intuitively, the brickwork pattern and the unitarity of propagation in time restricts
nonzero correlations between local operators to a light cone in space while
the unitarity of propagation in space restricts nonzero correlations to a light
cone in time.
Consequently, correlations can be non-vanishing only on the intersection of both light 
cones, namely on light-rays of the form $y=x \pm 2t$.
This intuitive picture has been made rigorous in Ref.~\cite{BerKosPro2019b} 
and has been found to be structurally stable upon generic, non-dual-unitary 
perturbations \cite{KosBerPro2021}, i.e., the dual-unitary contribution dominates
dynamical correlations even under small perturbations.
We briefly review the results obtained in Ref.~\cite{BerKosPro2019b} in what follows.

Consider a Hermitian and traceless operator
$a \in \text{End}(\mathbb{C}^q)$ and its 
embedding into $\text{End}(\Hil)$ for a fixed
lattice site $x \in \{0, \ldots, 2L-1\}$
given by
\begin{align}
a_x = \left(\bigotimes_{y=0}^{x-1}\mathds{1}_{\Hil_y}\right)\otimes a \otimes
\left(\bigotimes_{y=x+1}^{2L-1}\mathds{1}_{\Hil_y}\right).
\end{align}
That is, $a_x$ acts nontrivially only at lattice site $x$.
Note that being traceless ensures that $a$ and $a_x$ are Hilbert-Schmidt
orthogonal to the respective identities $\mathds{1}_{\mathbb{C}^q}$ and 
$\mathds{1}_{\Hil}$.
We will always assume $a$ to be Hilberg-Schmidt normalized, i.e., $\tr(a^2)=q$ and thus
$\tr(a_x^2)=q^{2L}$ for all $x$.
We are interested in dynamical correlation functions
$C_{ab}(x,y,t)  = \langle  a_x(t)b_y \rangle$
between such local operators $a_x$ and $b_y$, where 
$a_x(t)=\U^{-t}a_x\U^t$ is the time evolved operator in the Heisenberg picture.
The average is taken with respect to the 
infinite temperature state, $\langle \, \cdot \,  \rangle = q^{-2L}\tr(\,\cdot\,)$,
which constitutes the natural thermodynamic steady state for 
the quantum circuits under consideration.
Thus the dynamical correlations are given by
\begin{align}
    C_{ab}(x,y,t) = q^{-2L}\tr\left(\U^{-t}a_x\U^tb_y\right).
    \label{eq:cfdef}
\end{align}
For times $t \leq \lfloor L/2 \rfloor$ the intuitive picture sketched above applies.
More precisely a trivial extension of the results obtained in  Ref.~\cite{BerKosPro2019b}
to the present case of different local gates $U_1$ and $U_2$ in each half time step
yields
\begin{align}
   C_{ab}(x,y,t)  = \delta_{(x+ 2\nu t - y) \,(\mod 2L)}\frac{1}{q}\tr\left(\M_{\nu}^t(a) b\right).
   \label{eq:correlations_local_ops}
\end{align}
The first factor is a Kronecker delta which is one if $y = x + 2\nu t \, (\mod 2L)$ and zero otherwise,
where $\nu = -1 $ if $x$ is even and $\nu = 1$ if $x$ is odd.
Thus it restricts nonzero correlations to light rays $y = x + 2\nu t \, (\mod 2L)$ for initial times $t \leq \lfloor L/2 \rfloor$.
Due to our choice of periodic boundary conditions this statement applies to all lattice 
sites $x$ and we do not have to take scattering effects at open boundaries  into account, which would spoil
Eq.~\eqref{eq:correlations_local_ops} for $x$ close to the boundary.
In the second factor of the right hand side of Eq.~\eqref{eq:correlations_local_ops} the
trace is taken in $\text{End}(\mathbb{C}^q)$ and $\M_{\nu}$ is a
CPTP map $\text{End}(\mathbb{C}^q)\to \text{End}(\mathbb{C}^q)$.
Similar to the structure of the whole circuit $\M_{\nu}$ factors into two half time steps 
as $\M_{\nu}=\M_{\nu,2}\M_{\nu,1}$ with
\begin{align}
    \M_{+, i}(a) &= 
    \frac{1}{q}\tr_1\left[U_i^{-1} \left(a \otimes \mathds{1}_{\mathbb{C}^q}\right) U_i \right], \\
    \M_{-, i}(a) &= 
    \frac{1}{q}\tr_2\left[U_i^{-1} \left(\mathds{1}_{\mathbb{C}^q} \otimes a \right) U_i\right],
\end{align}
where the partial trace is taken over the first or second tensor factor of $\mathbb{C}^q \otimes\mathbb{C}^q$ respectively.

As a consequence the dynamics of correlation functions between local
operators is determined by the properties of the maps $\M_{\nu}$ in
the thermodynamic limit $L \to \infty$.
More precisely, the decay of such correlations is encoded in the
spectrum of $\M_{\nu}$, which is contained within the complex unit disk.
The spectral properties thus allow for a classification 
of dual-unitary circuits by their ergodic properties.
In any case $\M_{\nu}$ has the trivial eigenvalue $1$ with eigenvector
(eigenoperator) $\mathds{1}_{\mathbb{C}^q}$, meaning
that the maps $\M_{\nu}$ are also unital.
Moreover, the real subspace of traceless Hermitian operators of $\text{End}(\mathbb{C}^q)$
is invariant under $\M_{\nu}$.
Here, we consider the case where $\M_{\nu}$ is strictly contracting on this subspace, i.e., it has no
other eigenvalue of modulus one other than the trivial one.
This class of dual-unitary circuits is called ergodic and mixing as in the thermodynamic
limit all dynamical correlations $C_{ab}(x,y,t)$ tend
to zero as $\lambda_{\nu}^t$ for $t \to \infty$, where $\lambda_{\nu}$ is the largest (in modulus) nontrivial
eigenvalue of $\M_{\nu}$ \cite{BerKosPro2019b}.
Additionally, ergodic and mixing dual-unitary circuits exhibit quantum chaos in the spectral sense as their spectral form factor exactly follows random matrix predictions
\cite{BerKosPro2021:p}.
They moreover generate linear growth of operator entanglement 
and hence of the complexity of local operators upon time evolution \cite{BerKosPro2020a}.

\section{Spectral function for sums of local observables \label{sec:spectral_function}}

Having introduced the models under consideration we proceed by 
commenting on the spectral function $f(\omega)$ which appears in the formulation
of ETH in Eq.~\eqref{eq:ETH}.
To this end, consider a Hermitian and traceless operator $A \in \text{End}(\Hil)$.
We briefly discuss how the spectral function corresponding to $A$ emerges from 
statistical properties of its matrix elements as well as from its dynamical
autocorrelation function.
For the latter we derive our main result in the form of an asymptotic 
large $L$ expression
for the spectral function for a class operators, which
consists of spatial sums of the local operators discussed in the
previous section.

\subsection{Spectral function from the distribution of matrix elements
\label{sec:sprectral_func_from_distribution}}

In the following we describe the spectral function as the second moment of the frequency resolved
distribution of matrix elements of $A$ in the basis of eigenstates of $\U$.
We study the statistics of matrix elements $A_{mn}$ for which $\omega_{mn}$ is
sufficiently close to a given frequency $\omega \in \left[-\pi, \pi\right)$.
More precisely, we fix $0 < \Delta \ll 1$ and consider only matrix elements for which
$\omega_{mn} \in I_{\Delta}(\omega) := \left[\omega - \Delta/2, \omega + \Delta/2 \right)$.
The second moment, i.e., the variance of their distribution is given by
\begin{align}
    \text{var}_{\omega} (A_{mn}) = \frac{1}{N(\omega)}\sum_{m,n = 0}^{q^{2L}-1}|A_{mn}|^2 
    1_{I_{\Delta}(\omega)}(\omega_{mn})
    \label{eq:variance}
\end{align}
where  $1_{I_{\Delta}(\omega)}$ denotes the characteristic function
of $I_{\Delta}(\omega)$.
The normalization $N(\omega)$ is determined by the number of frequencies $\omega_{mn}$ in
$I_{\Delta}(\omega)$, i.e., 
\begin{align}
N(\omega) = \sum_{m,n = 0}^{q^{2L}-1}
1_{I_{\Delta}(\omega)}(\omega_{mn}).
\end{align}
Asymptotically one has $N(\omega) =  q^{4L}\Delta /(2\pi)$ as in chaotic systems the eigenphases 
and consequently also the frequencies are uniformly distributed in $\left(-\pi, \pi\right]$.
Note that  at $\omega=0$ the $q^{2L}$ real diagonal matrix elements contribute systematically to $N(0)$.
We neglect this issue as on the one hand we restrict most of our discussion to nonzero frequencies, and on the other hand, for $\Delta \gg 2\pi q^{-2L}$ the statistics 
of matrix elements is not significantly
affected by the diagonal matrix elements.

For operators $A$ compatible with the two-site shift invariance of $\U$, i.e. $\left[A, \T^2\right]=0$,
matrix elements between eigenstates of different momentum vanish.
Therefore, we may consider the distribution of matrix elements of the operator 
$\Proj_k A \Proj_k$ in the basis of eigenvectors of $\Proj_k \U \Proj_k$ corresponding to nonzero 
eigenvalues, where $\Proj_k$ denotes the orthogonal projection onto the eigenspace of $\T^2$ 
corresponding to momentum $k$.
We denote the frequency resolved second moment of this distribution by 
$\text{var}_{\omega}^{(k)}\!\left(A_{mn}\right)$, for which the normalization asymptotically reads
$N_k(\omega) = q^{4L}\Delta /(2\pi L^2)$ \cite{PinPro2007}, and obtain 
the variance of the full distribution by 
\begin{align}
\text{var}_{\omega}(A_{mn}) =  \frac{1}{L^2}\sum_{k=0}^{L-1}\text{var}_{\omega}^{(k)}\left(A_{mn}\right),
\label{eq:var_omega_k}
\end{align}
see App.~\ref{app:variance} for a derivation.

Independent from the symmetries of $A$, the spectral function can
be related to the variance of matrix elements $A_{mn}$ by replacing the latter  
with the ETH ansatz \eqref{eq:ETH} in Eq.~\eqref{eq:variance}.
Using the smoothness of $f(\omega)$ as well as the assumption that $R_{mn}$
has unit variance this yields
\begin{align}
|f(\omega)|^2 = q^{2L}\text{var}_{\omega} (A_{mn}).
 \label{eq:spectral_func_variance}
\end{align}
Thus, the spectral function is determined by the variance of the frequency
resolved distribution of
matrix elements.

\subsection{Spectral function from autocorrelation functions}

On the other hand, the spectral function may be extracted from the autocorrelation
function $\langle A(t) A \rangle$ of the operator $A$.
Representing $\U^{\pm t}$ by its respective spectral decomposition yields
\begin{align}
    g(t) := \langle A(t) A \rangle = q^{-2L}\sum_{m, n = 0}^{q^{2L}-1}|A_{mn}|^2 \ue^{\ui  \omega_{mn}t}
\end{align}
for the autocorrelation function.
Note that due to the cyclicity of the trace-like infinite temperature state one has $g(t)=g(-t)$.
Fourier transforming $g(t)$ as $ \hat{g}(\omega) = \sum_t g(t) \ue^{\ui \omega t}$ gives
\begin{align}
    \hat{g}(\omega) = 2\pi q^{-2L}\sum_{m, n = 0}^{q^{2L}-1}|A_{mn}|^2 \delta(\omega - \omega_{mn}).
\end{align}
This singular expression is regularized by integrating over $I_{\Delta}(\omega)$ giving the 
second moment~\eqref{eq:variance} which in combination with Eq.~\eqref{eq:spectral_func_variance}
leads to
\begin{align}
    |f(\omega)|^2 = \frac{q^{4L}}{2\pi N(\omega)}\int_{I_{\Delta}(\omega)} \hat{g}(\tilde{\omega})\ud \tilde{\omega}.
    \label{eq:spectral_function_integral}
\end{align}
Often evaluating the integral is omitted by considering autocorrelations only up to some
finite cut-off time $T$.
This effectively broadens the $\delta$ distributions as they are replaced by a strongly peaked
smooth function of effective width $\sim1/T$.
We denote the resulting finite time Fourier transform by $\hat{g}_T(\omega) = \sum_{t=-T}^T g(t) \ue^{\ui \omega t} $.
Choosing $\Delta$ sufficiently small compared to the scale on which $\hat{g}_T(\omega)$
varies we may replace the integral in Eq.~\eqref{eq:spectral_function_integral} by 
$\Delta \hat{g}_T(\omega)$ yielding 
\begin{align}
|f(\omega)|^2 = \hat{g}_{T}(\omega).
\end{align}
Thus the spectral function is just a regularized Fourier transform of 
an autocorrelation function of $A$.

\subsection{Asymptotical spectral functions \label{sec:asymptotic_spectral_function}}

Our objective in the following is to identify a class of operators for which
$\hat{g}_T(\omega)$ can be calculated exactly for sufficiently large cut-off times
$T$. 
The local Hermitian traceless operators $a_x$ discussed in Sec.~\ref{sec:circuits} seem as
a natural candidate but dual-unitarity causes their autocorrelation functions to vanish exactly for
$t < L$.
This may be overcome by considering sums of such operators acting on different
lattice sites instead.
The simplest and most convenient examples of such operators are sums
 \begin{align}
A = \frac{1}{\sqrt{L}}\sum_{x=0}^{L-1}a_{2x + \mu}, 
\label{eq:extensive_sum}
\end{align}
which act nontrivially on the even ($\mu = 0$) or on the odd ($\mu = 1$) sublattice only.
Moreover, $A$ is normalized as
$\langle A^2 \rangle = 1$.
This normalization ensures that the following results do not
depend on the system size $L$.

The autocorrelation function for $A$ can be decomposed into correlation functions
of the local operators $a_{2x + \mu}$ as
 \begin{align}
\langle A(t)A \rangle = \frac{1}{L}\sum_{x,y=0}^{L-1}C_{aa}(2x+\mu,2y+\mu,t)
\end{align}
For times $t  \leq \lfloor L/2 \rfloor$ the individual contributions
$C_{aa}(2x+\mu,2y+\mu,t)$ are given by Eq.~\eqref{eq:correlations_local_ops} and thus
are nonzero only if $y = x + \nu t \, (\mod L)$, where $\nu = 2\mu - 1$.
More precisely, we find
 \begin{align}
\langle A(t)A \rangle & = \frac{1}{L}\sum_{x=0}^{L-1}C_{aa}(2x+\mu,(2x+\mu + 2\nu t)\,(\mod 2L) ,t) \nonumber \\
& = \frac{1}{q}\tr\left(\M_{\nu}^t(a) a\right) \label{eq:auto_corr_A}
\end{align}
independent of the system size $L$.
Therefore we choose $T=\lfloor L/2 \rfloor$ as the cut-off time and obtain
 \begin{align}
\hat{g}_{\lfloor L/2 \rfloor}(\omega; a) = 1 + \frac{2}{q}\sum_{t=1}^{\lfloor L/2 \rfloor}
\cos(\omega t)\,\tr\left(\M_{\nu}^t(a) a\right),
\label{eq:g_omega}
\end{align}
where we explicitly include the dependence on $a$ in the notation.
As the expression~\eqref{eq:auto_corr_A} decays
exponentially at least as $\lambda_{\nu}^t$ for
$t \to \infty$, with $\lambda_{\nu}$ the largest modulus eigenvalue of $\M_{\nu}$, its
Fourier transform is well behaved for $L \to \infty$ and thus
Eq.~\eqref{eq:g_omega} gives rise to a
smooth function.
This determines the large $L$ asymptotics of the spectral function
\begin{align} 
   |f_{\infty}(\omega; a)|^2 = 1 + \frac{2}{q}\sum_{t=1}^{\infty}
\cos(\omega t)\,\tr\left(\M_{\nu}^t(a) a\right)
   \label{eq:spectral_function_asympt}
\end{align}
and constitutes our first main result.
As it is obtained for times up to $T = \lfloor L/2 \rfloor$ one can expect this to yield the
spectral function for  frequencies $|\omega| \gtrsim \pi/L$ in finite systems and thus 
for all nonzero frequencies in the thermodynamic limit $L  \to \infty$.
Deviations of $f(\omega; a)$ around $f_{\infty}(\omega; a)$ are due to the correlations
deviating from Eq.~\eqref{eq:auto_corr_A} for times $t > \lfloor L/2 \rfloor$.
More precisely, for thermalizing systems the autocorrelation
function equilibrates to a value 
exponentially small in system size, determined by the diagonal matrix elements 
corresponding to $\omega = 0$, and exhibits residual fluctuations only.
These times, however, do not contribute to the spectral function due to the smoothing 
within the finite frequency window $I_{\Delta}(\omega)$ of width $\Delta$ present in
Eqs.~\eqref{eq:variance}~and~\eqref{eq:spectral_function_integral}, respectively.
In contrast, deviations from the asymptotic spectral function
at finite system size may be due to the
the dynamics of autocorrelation functions at intermediate time scales.
We postpone the discussion thereof to the next section when considering a concrete
example.

\begin{figure}[]
    \includegraphics{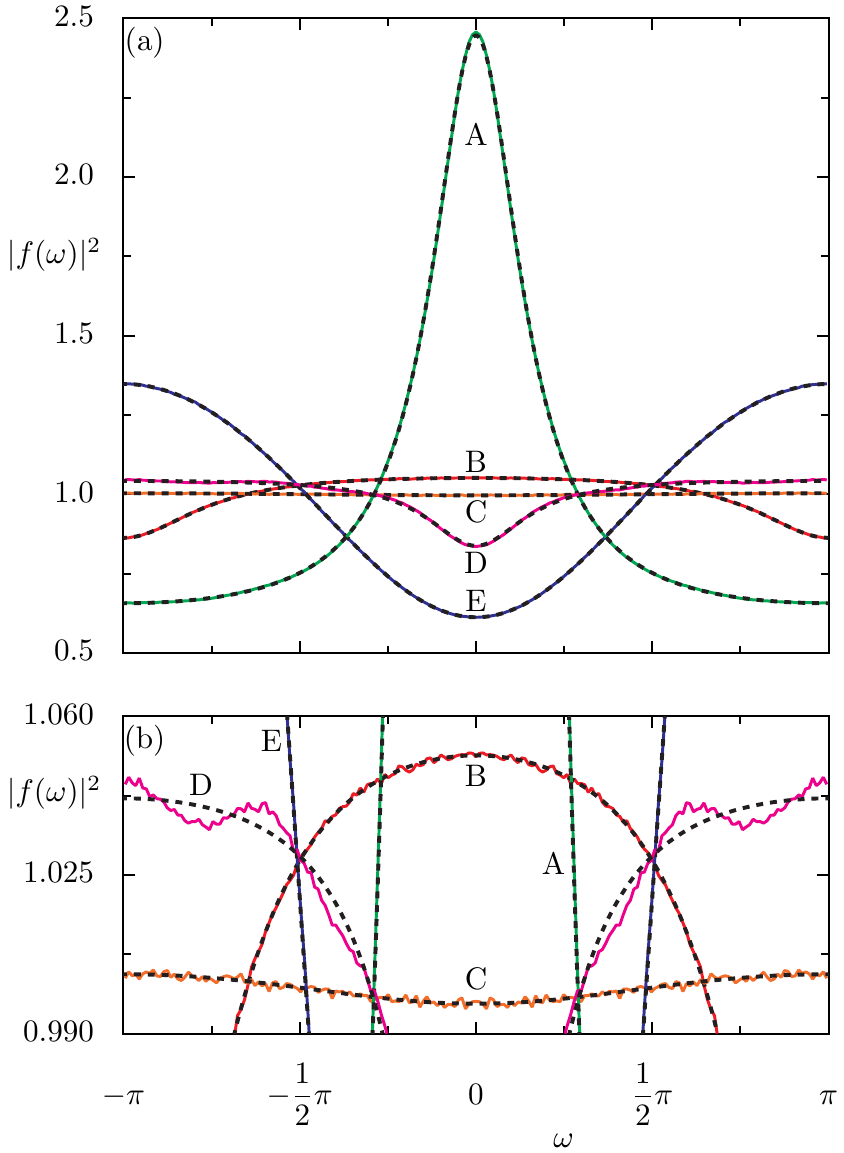}
    \caption{Spectral function of the operator $A$, Eq.\eqref{eq:extensive_sum} for five representative realizations of the dual-unitary circuit $\U$ with $2L=16$ labeled by A-E. The corresponding asymptotic expressions~\eqref{eq:spectral_function_sigmaZ} are depicted as dashed black lines. The lower panel (b) is a magnification of panel (a).
    }
    \label{fig:spectral_func}
\end{figure}

A detailed analysis of $|f_{\infty}(\omega; a)|^2$ is possible
for a suitable choice of the operator $a$.
To this end let $a=a_{\lambda}$ be a Hermitian eigenvector of $\M_{\nu}$
corresponding to a real eigenvalue $\lambda$.
We may write $|\lambda_{\nu}|=\ue^{-\gamma}$ with $\gamma > 0$.
The latter is the rate with which the autocorrelation Eq.~\eqref{eq:auto_corr_A} decays, i.e.,
$\langle A(t)A \rangle = \lambda^t \propto \ue^{-\gamma t}$.
This allows for evaluating its Fourier transform exactly, see App.~\ref{app:fourier_transfo}, yielding
 \begin{align}
 |f_{\infty}(\omega; a_{\lambda})|^2 = \frac{\sinh(\gamma)}{\cosh(\gamma) - \text{sign}(\lambda) \cos(\omega)}.
  \label{eq:f_analytic}
\end{align}
For fast decaying modes, i.e., large $\gamma$ the spectral function~\eqref{eq:f_analytic} is essentially flat, $|f_{\infty}(\omega; a_{\lambda})|^2 \approx 1$.
In contrast, for slowly decaying modes, i.e., $\gamma \ll 1$ the spectral function is strongly peaked around $\omega=0$ ($\lambda>0$) or $\omega = \pi$ ($\lambda<0$).
Indeed, by expanding both numerator and denominator up to second order in $\gamma$ around $0$ and
$\omega$ around the peak yields a Lorentzian spectral function with full width at
half maximum given by $2\gamma$ and peak height $2/\gamma$;
see Fig.~\ref{fig:spectral_func}(a) for some examples of the shape of
$|f_{\infty}(\omega; a_{\lambda})|^2$.
Note, that the (leading) eigenvalues of $\M_{\nu}$ are not necessarily real, but may come in complex conjugate 
pairs $\lambda, \bar{\lambda} \in \mathbb{C}$, for which the corresponding eigenvectors are not Hermitian
but come in Hermitian conjugate pairs.
Moreover, the linear map $\M_{\nu}$ is in general not diagonalizable leading to possibly nontrivial 
Jordan blocks (in case of degenerate eigenvalues) and thus to polynomial corrections to the pure
exponential decay of correlations when considering
generalized eigenvectors.

\section{Numerical tests \label{sec:numerics}}

In this section we compare the asymptotic
expressions, Eqs.~\eqref{eq:spectral_function_asympt}~and~\eqref{eq:f_analytic},
with spectral functions obtained from exact diagonalization of representative example
systems, for which we find excellent agreement in generic cases.
Additionally, we study the frequency resolved distribution of matrix elements numerically
in more detail.
In particular, we confirm that the latter coincides with 
a complex Gaussian distribution for frequencies $|\omega|>\pi/L$ by computing
higher moments of the distribution.

For our numerical simulation we consider chains of qubits, i.e., $q=2$ for which
an exhaustive parameterization of dual-unitary gates $U \in \text{U}(4)$ 
is given by
\cite{BerKosPro2019b}(SM)
\begin{align}
U =  \left(u_+ \otimes u_- \right)V(J) \left(v_- \otimes v_+ \right).
\label{eq:gates_parametrisation}
\end{align} 
Here, $u_{\pm}, v_{\pm} \in \text{U}(2)$ and
\begin{align}
V(J) = 
\exp\left(-\ui\left[ \frac{\pi}{4} \sigma^x \otimes \sigma^x  + 
\frac{\pi}{4}  \sigma^y \otimes \sigma^y  + J  \sigma^z \otimes \sigma^z \right]\right)
\end{align} 
with $J \in \left(\pi/4, \pi/4 \right]$ and Pauli matrices $\sigma^x$, $\sigma^y$, and $\sigma^z$.
In the following we set $J=0$ in both $U_1$ and $U_2$
rendering the resulting circuits ergodic and mixing \cite{BerKosPro2019b}(SM).
Moreover, for the corresponding local $\text{U}(2)$ matrices, denoted 
by $u_{\pm, i}, v_{\pm, i}$ for both half time steps we choose generic examples reported in Appendix~\ref{app:parameters}.
We consider the operator $a=\sigma^z \in \text{End}(\mathbb{C}^2)$
in order to construct $A$ via Eq.~\eqref{eq:extensive_sum} on the 
odd sublattice ($\mu=\nu=+1$).
We fix the lattice size as $2L = 16$ in what follows.

\begin{figure}[bh]
    \includegraphics{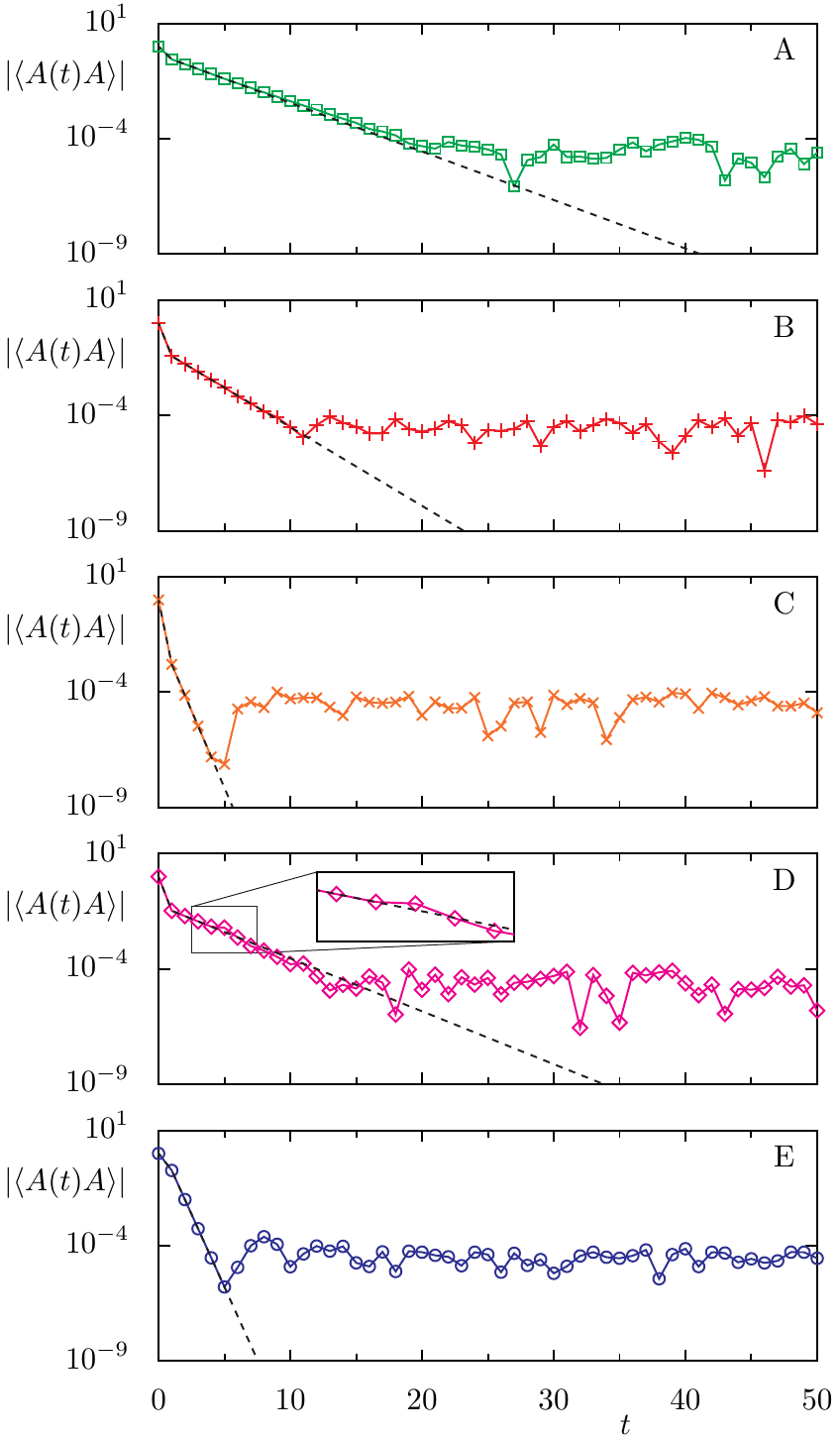}
    \caption{Autocorrelation functions of the operator $A$ versus time for systems A-E 
        with $2L=16$ 
        on a semilogarithmic
        scale (colored symbols connected by lines). The asymptotic autocorrelation 
        functions $\alpha \lambda_{\nu}^t$ ($t \geq 1$) are depicted as black dashed lines.
    }
    \label{fig:autocorr}
\end{figure}

\subsection{Spectral function}

For the quantum circuits introduced above we compute the spectral function of $A$ as
 described in Sec.~\ref{sec:sprectral_func_from_distribution}
utilizing the two-site shift invariance of $A$, i.e., $\left[\T^2, A \right]=0$.
We first construct the projectors $\Proj_k$ onto the subspaces of momentum $k$ and subsequently diagonalize $\U$ in each subspace.
This allows for computing the variance $\text{var}_{\omega}^{(k)}(A_{mn})$ for each momentum $k$.
To this end we subdivide the interval $\left[-\pi, \pi\right)$ into $N_{\Delta}=201$ subintervals
$I_{\Delta}(\omega_i)$ of width $\Delta = 2\pi/N_{\Delta}$ with centers $\omega_i = 2\pi (i + 1/2)/N_{\Delta} - \pi$ for $i \in \{0, \ldots, N_{\Delta}-1\}$.
Having determined $\text{var}_{\omega}^{(k)}(A_{mn})$ we obtain $\text{var}_{\omega}(A_{mn})$ from 
Eq.~\eqref{eq:var_omega_k}.
Eventually this yields the spectral function via 
Eq.~\eqref{eq:spectral_func_variance}.
We obtain the spectral function by this procedure
for five different realizations of the circuit labeled by A-E as depicted
in Fig.~\ref{fig:spectral_func} by thin (colored) lines.
The spectral functions vary between being
almost flat (System C) and being sharply peaked (System A) therefore covering the typical behavior
predicted by Eq.~\eqref{eq:f_analytic}.
Figure~\ref{fig:spectral_func}(b) depicts a magnification around $|f(\omega)|=1$.

As $a=\sigma^z$ is not an eigenvector of $\M_{\nu}$ in general the description
presented in Sec.~\ref{sec:asymptotic_spectral_function}
does not apply directly.
Nevertheless,  the latter can be adapted to the qubit circuits at
hand, which we illustrate in the following.
We start by noting that in the present case $\M_{\nu}$ can 
generically be diagonalized 
and for $J=0$ exhibits two eigenvalues equal to $0$ and only one nontrivial and nonzero
eigenvalue $\lambda$, which is necessarily real.
We report these eigenvalues for both $\M_{+}$ and $\M_{-}$ in Table~\ref{tab:eigenvalues} for the
systems under consideration.
We denote the corresponding hermitian eigenvector by $a_{\lambda}$.
As $\M_{\nu}$ can be diagonalized we can expand $a$ in the basis consisting of its 
nonorthogonal eigenvectors.
To this end let 
$\bar{a}_{\lambda}$ denote the hermitian and traceless 
left eigenvector corresponding to $\lambda$.
That is $\bar{a}_{\lambda}$ is an eigenvector with eigenvalue $\lambda$
of the adjoint of $\M_{\nu}$ with respect to the 
Hilbert-Schmidt scalar product on $\text{End}(\mathbb{C}^2)$.
We define
$\alpha := \tr(\bar{a}_{\lambda} a)\tr(a_{\lambda} a)/(q\tr(\bar{a}_{\lambda} a_{\lambda}))$,
for which one has $|\alpha|\leq 1/|\lambda|$,
see Table~\ref{tab:eigenvalues} for the numerical values in
the systems considered here.
The above definition allows for writing the autocorrelation functions 
$\langle A(t)A \rangle = \tr\left(\M_{\nu}^t(a) a\right)/q = \alpha \lambda_{\nu}^t$ for $1 \leq t \leq \lfloor L/2 \rfloor$.
This is illustrated in Fig.~\ref{fig:autocorr} for the systems A-E where the numerically 
obtained initial dynamics
of autocorrelation functions is depicted for $2L=16$ by colored symbols.
For times up to $t = \lfloor L/2 \rfloor = 4$ they follow the exponential decay described above
and depicted by dashed black lines.
For slowly decaying modes as in the systems A, B, and E the exponential decay proportional to
$\lambda_{\nu}^t$ approximately continues also for times larger than $\lfloor L/2 \rfloor$.
At large times autocorrelation functions are expected to equilibrate, i.e., 
they oscillate around their long-time average
with both the oscillations and the equilibrium value of order $q^{-2L}$. 
For the systems A-E this behavior is approached already on the timescale shown here.
More precisely, the relaxation time $T^{*}$ until the 
equilibrium behavior is reached can be
roughly estimated by the time $T^{*}_{\nu}$ for which $\lambda_{\nu}^{T^{*}_{\nu}}$ is of the order of $q^{-2L}$,
i.e., $T^{*}_{\nu} = -2L \ln(q)/\ln(|\lambda_{\nu}|)$.
For fast decaying modes, however, this may give $T^{*}_{\nu}<\lfloor L/2 \rfloor$
whereas the exponential decay
continues up to $t=\lfloor L/2 \rfloor$.
Note further that for times larger than $t=\lfloor L/2 \rfloor$
the dynamics of correlation 
functions in the two directions $\nu=+1$ and $\nu = -1$ is no longer independent from 
each other.
Thus an estimate for the equilibration time is given by
$T^{*}=\max\{\lfloor L/2 \rfloor, T^{*}_{+}, T^{*}_{-}\}$, taking into account the 
slowest possible decay of correlations in both directions as well as the initial 
dynamics for times $t\leq\lfloor L/2 \rfloor$. 
In general on has $T^* = \lfloor L/2 \rfloor$ if 
all eigenvalues $\lambda$
of $\M_{+}$ and $\M_{-}$ are bounded as $|\lambda| \leq q^{-4}$.
This is the case for the systems C and E, while we find $T^*=T^*_+$ 
for the systems A, B and D.

\begin{table}[]
    \caption{Nontrivial eigenvalues of the CPTP maps $\M_{\nu}$ and coefficients $\alpha$ for
    systems A-F.}
    \label{tab:eigenvalues}
    \centering
\begin{tabular}{|c|l|l|l|l|l|l|}
    \hline
    System & \hspace*{0.22cm} A & \hspace*{0.22cm}  B & \hspace*{0.22cm} C & 
    \hspace*{0.22cm}  D & \hspace*{0.22cm}  E & \hspace*{0.22cm}  F \\
    \hline 
    $\lambda_{+}$ & \textcolor{white}{-}0.616 & -0.455 & \textcolor{white}{-}0.046 &
     \textcolor{white}{-}0.590 & \textcolor{white}{-}0.055 & -0.225  \\
    \hline 
    $\lambda_{-}$ & \textcolor{white}{-}0.277 & -0.091 & -0.018 & \textcolor{white}{-}0.162 &
     \textcolor{white}{-}0.018 & \textcolor{white}{-}0.875  \\
    \hline 
    $\alpha$ & \textcolor{white}{-}0.450 & -0.082 & -0.035 &  -0.056 & -3.357 & -0.268  \\
    \hline
\end{tabular}
\end{table}

In any case, in order to obtain the asymptotic form of the spectral function, the Fourier transform~\eqref{eq:g_omega} can be evaluated exactly in the limit
$L \to \infty$ yielding
 \begin{align}
|f_{\infty}(\omega; a)|^2 = (1-\alpha) + \alpha|f_{\infty}(\omega; a_{\lambda})|^2,
\label{eq:spectral_function_sigmaZ}
\end{align}
where we used $\tr\left(\M_{\nu}^t(a) a\right)/q = \alpha \lambda_{\nu}^t$ for $t\geq 1$ 
as derived above.
Thus, the asymptotic spectral functions are composed of a flat background given by $(1-\alpha)$
due to the operators in the kernel of $\M_{+}$,
i.e., those operators for which dynamical correlations decay to $0$ within a single time step,
as well as the second term  $\alpha|f_{\infty}(\omega; a_{\lambda})|^2$ due 
to the slowest decaying mode. 
The resulting asymptotic spectral functions~\eqref{eq:spectral_function_sigmaZ}
for the systems A-E are depicted in
Fig.~\ref{fig:spectral_func}(a) as black dashed lines.
On the shown scale we find excellent agreement between the numerically obtained spectral
functions and their asymptotic counterpart.
Surprisingly, the asymptotic result agrees well with the numerically obtained
spectral functions even for small frequencies $0 < |\omega|\lesssim \pi/L$.

For slowly decaying autocorrelations, for which $|f_{\infty}(\omega, a)|^2$ is 
sharply peaked one has $|f_{\infty}(\omega, a)|^2 <  2\ue^{\gamma}/\gamma + 
\mathcal{O}(1)$, minding that $|\lambda| = \ue^{-\gamma}$.
Consequently, the observed agreement of the spectral functions at finite system size with the asymptotic expressions implies the variance of matrix elements to be bounded by 
$q^{-2L} 2\ue^{\gamma}/\gamma$ for nonzero frequencies in the present case.
In particular, they are exponentially small in system size up to a factor which is 
determined by the local gates $U_1$ and $U_2$ only and which does not depend on the 
operator $a$.
Note that by fine tuning the system the above bound might still become arbitrary large.

\subsection{Finite size deviations}

\begin{figure}[t]
    \includegraphics{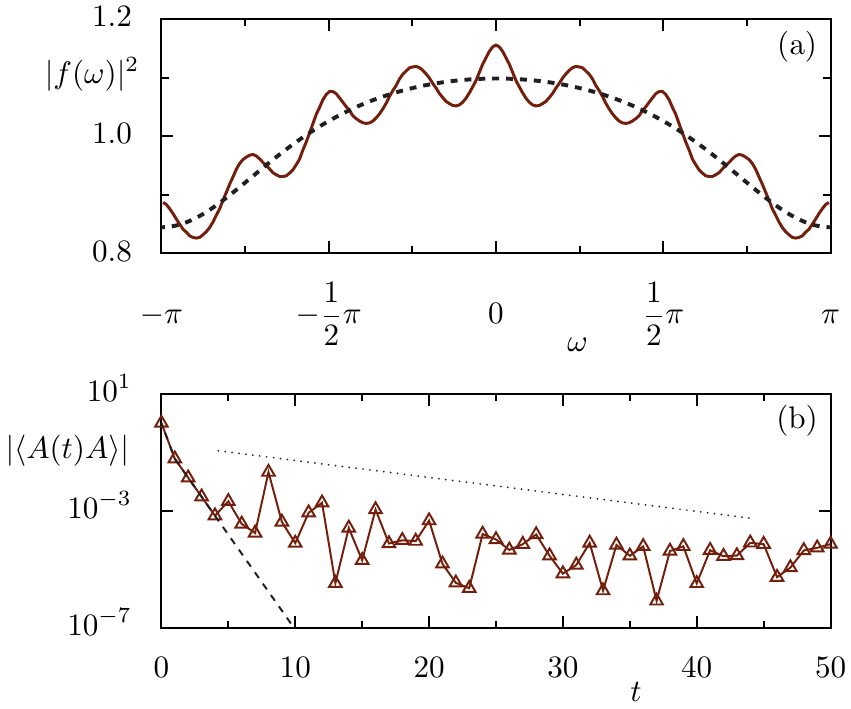}
    \caption{System F
        at system size 
        $2L=16$. Panel (a) depicts the numerically obtained (thin brown line) and the 
        asymptotic (black dashed line) spectral function. Panel (b) shows the corresponding 
        autocorrelation functions (brown triangles connected by a thin line) and the asymptotic 
        autocorrelation functions $\alpha \lambda_{+}^t$ ($t \geq 1$) (black dashed line). The dotted 
        black line indicates the asymptotically
        slowest possible decay proportional to $\lambda_{-}^t$.
    \label{fig:failure}}
    \vspace*{0.5cm}
    \includegraphics{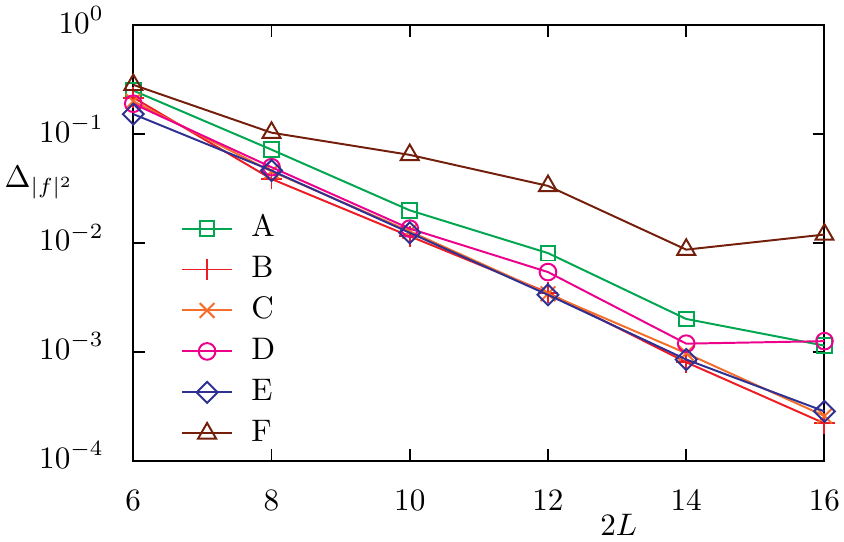}
    \caption{Difference $\Delta_{|f|^2}$ of the spectral function and its asymptotic value
        versus system size $L$. The lines connecting the symbols serve as a guide to the eye.
    \label{fig:diff}}
\end{figure}

Despite the good agreement between the numerically obtained spectral function with the 
asymptotic expression we find deviations at finite system size.
These deviations become visible in Fig.~\ref{fig:spectral_func}(b), which shows a
magnification of panel (a), and are particularly evident for system D.
We analyse such finite size deviations in the following in more detail by relating them 
to the dynamics of autocorrelation functions at intermediate time scales $\lfloor L/2 
\rfloor < t < T^*$, i.e., when $T^*>L/2$.
At these times Eq.~\eqref{eq:auto_corr_A} is not valid and deviations from the 
exponential decay of autocorrelation functions proportional to $\lambda_{\nu}^t$ are possible.
That is, there might be a time $t^*$ at which $\langle A(t^*)A\rangle \gg \lambda_{\nu}^{t^*}$.
Upon Fourier transform this contributes to the spectral function as 
$2\langle A(t^*)A\rangle \cos(\omega t^*)$ leading to oscillations of the spectral 
function  around the asymptotic result proportional to $\cos(\omega t^*)$.
Indeed for system D one finds such a behavior at time $t^*=5$ as indicated by the inset 
for system D in Fig.~\ref{fig:autocorr}.
This explains small oscillations of the numerically obtained spectral function around its 
asymptotic counterpart.
However, this behavior will vanish as soon as $L > 2t^*$ or is at least shifted to later 
times and thus would be exponentially suppressed.
When studying the spectral function for system D at smaller system sizes (not 
shown) we find that the behavior described above is not as pronounced as at $2L=16$ and 
thus may appear for individual values of the system size only.
One therefore may ask if the
deviations described above can occur systematically.

In fact, we expect systematic deviations when autocorrelations in the opposite direction $-\nu$ decay 
much slower than in the initial direction $\nu$ as both directions are no longer 
independent at times $t > L/2$.
This corresponds to $|\lambda_{-\nu}|>|\lambda_{\nu}|$, where for the operators investigated here $\nu = +1$.
For such a situation, denoted as system F, we report these eigenvalues as well as the
coefficient $\alpha$ in Table~\ref{tab:eigenvalues}.
Both the spectral function and the autocorrelation function is depicted in 
Fig.~\ref{fig:failure}(a)~and~(b), respectively.
Most prominently, one finds $\langle A(t^*)A \rangle \gg \lambda_{\nu}^{t^*}$
at time $t^*=8$
which leads to significant oscillations of the numerically obtained spectral function around the corresponding asymptotic result.
A heuristic argument for the occurrence of such phenomena can be given as follows.
In the previous section we saw that for slowly decaying autocorrelations the exponential
decay proportional to $\lambda_{\nu}^t$ approximately continues for times $t > L/2$.
Thus in the situation at hand there might be contributions to the autocorrelation function
decaying as $\lambda_{-\nu}^t$ and thus much slower than $\lambda_{\nu}^t$.
Ultimately this contribution might dominate the autocorrelation function at some time $t^*$ as it is observed in system F.
Nevertheless, these contribution should be of the order of 
$|\lambda_{-\nu}|^{t^*} \leq|\lambda_{-\nu}|^{L/2}$ and therefore vanish exponentially with increasing system size as well.

For the numerically accessible system sizes we quantify the deviations of numerically 
obtained and asymptotic spectral function as well as their scaling with system size.
We choose their $\text{L}^2$ distance 
$\Delta_{|f|^2}:=\big\| |f|^2 - |f_{\infty}|^2 \big\|_2$ as a measure for the deviations.
This is shown in Fig.~\ref{fig:diff} for system sizes $L \in \{3, \ldots, 8\}$.
For the systems A-E we find these differences to exponentially decrease with
increasing system size at an approximately equal rate for the
system sizes accessible by exact diagonalization.
As discussed above deviations are accidentally larger at $2L=16$  for system D and thus 
$\Delta_{|f|^2}$ does not follow the overall exponential decay at this system size.
System A shows a slightly slower decay as well as slightly larger deviations.
The latter are dominated by small frequencies $|\omega|\lesssim\pi/L$.
For system F, as expected, $\Delta_{|f|^2}$ is much larger than for the other systems.
Although there is an overall decrease of the deviations with increasing
system size, the
numerically accessible system sizes are not sufficient to extract the exact scaling and 
and to ultimately confirm the conjectured exponential decay.

\subsection{Distribution of matrix elements and higher moments}

\begin{figure}[]
    \includegraphics{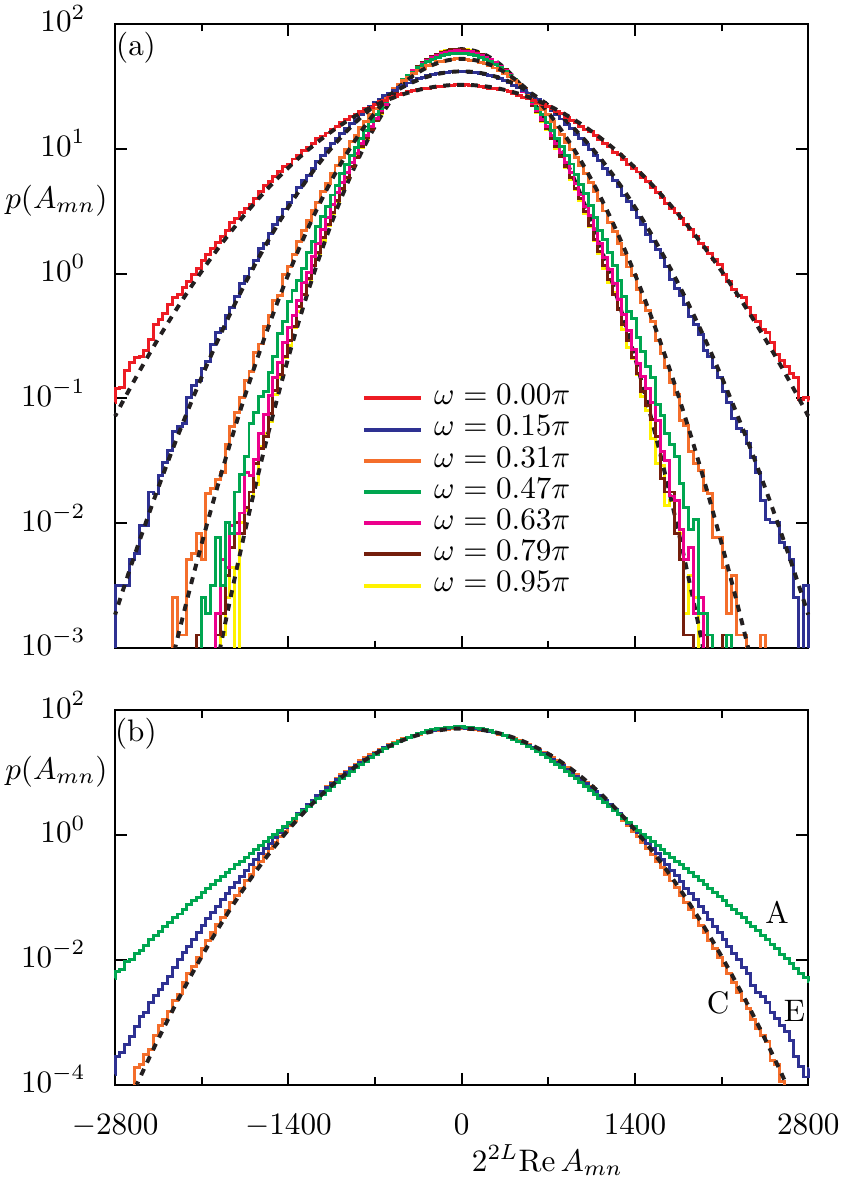}
    \caption{Distribution of real parts of matrix elements for $2L=16$.
        In panel (a) the distribution for system A at various frequencies (see legend) is shown while (b) depicts distribution of at all frequencies in the system A, C, and D. The black dashed lines show
        a Gaussian distributions corresponding to the variance determined by (a) $|f(\omega)|^2$ at $\omega \in \{0, 0.15\pi, 0.31\pi, 0.95\pi\}$ and (b)
        $|f(\omega)|^2=1$
    }
    \label{fig:distributions}
\end{figure}
\begin{figure}[]
    \includegraphics{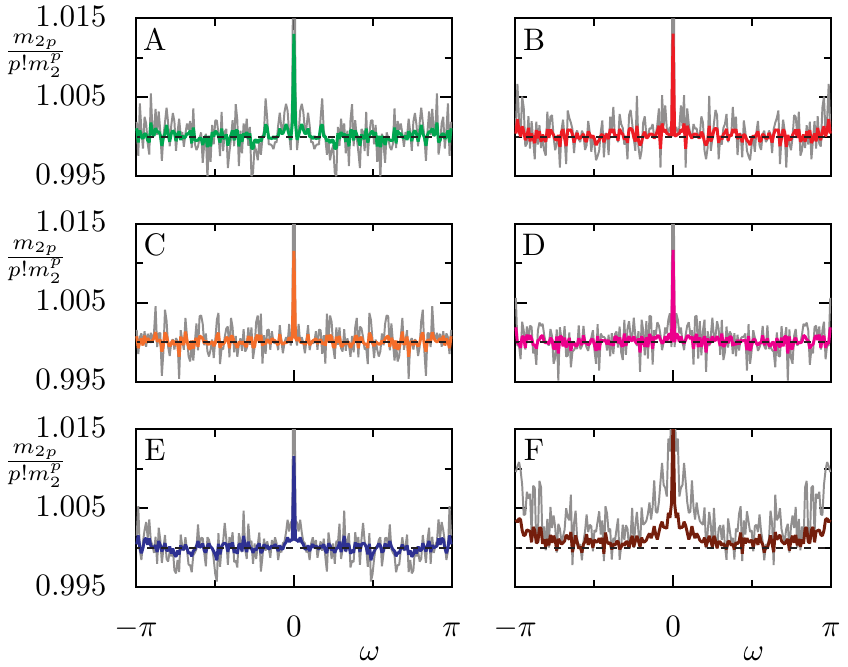}
    \caption{Rescaled fourth (thick colored line) and sixth (thin gray line) moment of the distribution of matrix elements for systems A-F with $2L=16$.
        The dashed black line corresponds to a Gaussian distribution.
    }
    \label{fig:moment4}
\end{figure}

Additionally, we investigate the distribution of matrix elements in systems A-F
numerically beyond the second moment.
To this end we depict the distribution of real parts of matrix elements of $A$ for 
various frequencies
for system A in Fig.~\ref{fig:distributions}(a).
Here, we explicitly choose system A as the sharply peaked spectral function indicates that
the differences in the distributions are most prominent compared to the systems B-F.
Indeed one finds a clearly broader distribution at small frequencies compared to larger
frequencies.
However, the shape of the distribution for nonzero frequencies is well described by a
Gaussian distribution with zero mean represented by dashed lines.
At $\omega=0$ the tails of the empirical distribution are slightly underestimated by the 
corresponding Gaussian distribution.
Note that we depict the distribution of matrix elements of $\Proj_k A \Proj_k$ combined for all
$k \in \{0, \ldots, L-1 \}$ rather than showing the distribution for
a fixed $k$ as it increases the number of matrix elements entering the statistics.
For this kind of sampling the variance of the Gaussian distribution 
is given by $q^{-2L}|f(\omega)|^2/(2L)$ as it is the marginal distribution of the 
complex Gaussian distribution of complex matrix elements
with twice that variance.
For nonzero frequencies the distribution of the imaginary parts of matrix elements (not shown)
coincides with that of the real parts.
For $\omega=0$, however, one finds an additional peak at $\text{Im}\,A_{mn}=0$ originating from the real
diagonal matrix elements.

Moreover, we depict the distribution of real parts of matrix elements not resolved by frequency in
Fig.~\ref{fig:distributions}(b) for the systems A, C, and D.
We do not include systems, B, E, and F as they can not be distinguished from system C on 
the shown scale.
For the almost flat spectral function of system C, $|f(\omega)|^2=1$, the resulting distribution is well 
described by the a Gaussian distribution with variance as discussed above (black dashed line).
In contrast for the sharply peaked spectral function of system A the distribution exhibits 
exponential tails from combining Gaussian distributions with different variances.
This deviation from a Gaussian is less pronounced but still visible for system D due to the 
corresponding spectral function being not flat.
Thus, as a non flat spectral function is due the presence of slowly decaying modes, the 
latter provide an explanation for heavy tails of the distribution of matrix elements, 
which was observed, for instance, in Ref.~\cite{RoyLevLui2018} as well.

While the spectral function, i.e., the second moment of the frequency resolved distributions
can be computed from the dynamics up to times proportional to the system size, higher moments can not
easily be obtained from dynamics up to this time scale.
We therefore study higher moments of even order
\begin{align}
    m_{2p}(\omega)  = \frac{1}{N(\omega)}\sum_{m,n = 0}^{q^{2L}-1}|A_{mn}|^{2p}
    1_{I_{\Delta}(\omega)}(\omega_{mn})
    \label{eq:moments}
\end{align}
numerically.
Note that Eq.~\eqref{eq:moments} 
coincides with the variance, Eq.~\eqref{eq:variance}, for $p=1$.
For a complex Gaussian distribution one has $m_{2p}(\omega)=p\text{!} m_2(\omega)^p$.
Using again the two-site shift invariance of $A$ we compute higher moments
of the distribution similarly to the variance.
We depict the rescaled moments $m_{2p}(\omega)/(p\text{!} m_2(\omega)^p)$ obtained by this procedure
for system A-F in
Fig.~\ref{fig:moment4} for $p=2$ (thick colored lines) and $p=3$ (thin gray lines) respectively.
Up to small fluctuations they reproduce the Gaussian value of one well for frequencies
$|\omega|\gtrsim\pi/L$ further confirming the Gaussian shape of the distribution of matrix elements.
Slight deviations appear for small frequencies and partially around $\omega = \pm \pi$.
Interestingly the deviations are largest for system F which we cannot readily 
explain by the dynamics of autocorrelation functions.
Note that an accurate estimation of higher moments requires larger and larger
system sizes and is not attempted here.

\section{Conclusions  \label{sec:conclusions}}

We study statistics of matrix elements for a class of operators
in dual-unitary quantum circuits.
For these operators we derive the asymptotic, 
large $L$, form of the spectral function, i.e., 
the second moment of the
frequency resolved distribution of their matrix elements.
This is achieved by utilizing the space-time dual-unitarity of the considered
quantum circuits in order to compute the exact short time dynamics
of autocorrelation functions.
The latter is determined by the local gates in terms of the CPTP maps $\M_{\nu}$.
Consequently, their spectrum allows for characterizing the asymptotic properties
of spectral functions.
More precisely, the presence of slowly decaying modes corresponding to large eigenvalues of
$\M_{\nu}$ may lead to sharply peaked spectral functions, 
whose peak height is inversely proportional
to the decay rate of theses modes independent of system size.
In contrast, the absence of slowly decaying modes will render
the spectral function essentially flat and structureless.
In any case, this confirms that the distribution of matrix elements has a variance which is
exponentially small in system size as it is implied by the ETH ansatz~\eqref{eq:ETH}. 

Comparing the asymptotic results with spectral functions obtained from exact diagonalization of 
one dimensional chains of qubits we find excellent agreement for generic dual unitary systems.
Additionally, by studying the initial dynamics of autocorrelation functions we confirm the
correspondence between the structure of spectral functions and the rate of relaxation of 
correlations
towards their equilibrium value.
This reveals the origin of the deviations between numerically obtained spectral functions and its asymptotic form due to the autocorrelation functions at intermediate times.
Exact diagonalization also allows for 
studying the frequency resolved distribution of matrix elements in more detail.
In particular, by computing higher moments of the distribution we confirm that the latter is
well described by a Gaussian. 
When disregarding the frequency dependence we find the combined distribution to deviate from 
a Gaussian distribution.
That is, the distribution exhibits exponential tails if the spectral function of the 
corresponding operator is not flat due to slowly decaying modes. 

We emphasize that the computation of the asymptotic spectral function from
exact dynamical correlations for initial times via the CPTP maps $\M_{\nu}$
can be straightforwardly extended to a larger class of operators built from
extensive sums of local operators.
For example, this includes operators supported on both the even and the odd
sublattice as well as operators, which violate translational 
invariance by choosing different nontrivial local
operators at different sites.
In principle also inhomogeneous dual-unitary quantum circuits in which the local
gates depend on the lattice site
can be treated similarly.
In the latter cases, however, the characterization of
spectral functions by the spectrum of a single
CPTP map $\M_{\nu}$ can no longer be applied directly.

Although the construction of the asymptotic spectral function presented here agrees well with 
numerical results in many cases, there may be deviations as discussed above.
These deviations are due to autocorrelation functions fluctuating around the exponential decay at intermediate times.
Controlling those fluctuations, i.e. rigorously proving that the spectral functions converge
towards the asymptotic result, 
requires knowledge of the
dynamics of autocorrelation functions for much larger times, i.e. up to Heisenberg time.
This would allow for exactly computing also higher moments of the distribution 
of matrix elements in order to confirm their Gaussian shape.
Computing dynamical correlations at large times exactly is currently out of the scope of 
the methods based on dual-unitarity presented here and is an important challenge for future work.
In contrast, the question whether the asymptotic form of the spectral function gives an accurate
description also for generic, i.e., nondual-unitary Floquet circuits may be checked within the 
perturbative framework of Ref.~\cite{KosBerPro2021}.

\vspace*{0.3cm}
\acknowledgments

We acknowledge fruitful discussion with D. Luitz as well as discussions and collaboration on related 
projects with B. Bertini and P. Kos.
The work has been supported by European Research Council (ERC) Advanced grant 694544-OMNES and 
Slovenian research agency (ARRS) research program P1-0402. FF further acknowledges support
by the Deutsche Forschungsgemeinschaft (DFG) -- 453812159 and thanks 
A. B{\"a}cker for providing PyxGraph for the creation of figures.

\appendix
%

\begin{table}[h!]
    \caption{Parameters of the local gates for systems A-F.}
    \label{tab:local_unitaries}
    \centering
    \begin{tabular}{|l| l|}
        \hline
        &  $\qquad \qquad \quad \quad \quad u_{\pm, i}$ and $v_{\pm, i}$ \\
        \hline
        A 
        &  $u_{+, 1}$ = \Uloc{-0.03440376+}{0.95921874}{0.10038637+}{0.26198924}{
            0.21897117+}{0.17540641}{-0.92413009-}{0.25936037} \\
        & $u_{-, 1}$ = \Uloc{0.79008463+}{0.42675293}{-0.3923305-}{0.19931128}{
            0.43912594-}{0.02857657}{0.89432201-}{0.08086989} \\
        & $v_{+, 1}$ = \Uloc{-0.48395602-}{0.26977298}{-0.38316526+}{0.73904905}{
            0.83244561-}{0.00658885}{0.02112852+}{0.55366459}\\
        & $v_{-, 1}$ = \Uloc{0.93026558-}{0.19510379}{-0.06303577+}{0.30424817}{
            0.25892305-}{0.17175364}{-0.16517923-}{0.93604239} \\
        & $u_{+, 2}$ = \Uloc{0.59082918+}{0.75484878}{-0.25218615-}{0.13238709}{
            0.07482491-}{0.27481891}{-0.15030173-}{0.9467234}  \\
        & $u_{-, 2}$ = \Uloc{-0.09222418+}{0.39020893}{0.68968198-}{0.60296804}{
            -0.17004669+}{0.90017543}{-0.28957816+}{0.27733157} \\
        & $v_{+, 2}$ = \Uloc{-0.04850466+}{0.19938165}{0.71066198+}{0.67294413}{
            0.67865799-}{0.70520748}{0.05770763+}{0.19691514} \\
        & $v_{-, 2}$ = \Uloc{0.79745641+}{0.20004401}{-0.09160851-}{0.56183054}{
            0.54323348+}{0.1701266}{0.08528023+}{0.81772955} \\
        \hline
        B 
        & $u_{+, 1}$ =  \Uloc{0.62703598+}{0.33406795}{0.07672879-}{0.69952639}{    -0.20388687+}{0.67353888}{-0.70782966+}{0.06125921} \\
        & $u_{-, 1}$ = \Uloc{0.16747675-}{0.16925005}{0.08962712-}{0.9670951}{
            0.76752061+}{0.59516222}{-0.23801086+}{0.00669881j}\\
        & $v_{+, 1}$ = \Uloc{0.7677496-}{0.2283061}{-0.30161091+}{
            0.51717281}{-0.10473253-}{0.58946412}{-0.6418985-}{0.47909233} \\
        &  $v_{-, 1}$ = \Uloc{-0.12026078-}{0.76763149}{-0.0356572+}{
            0.62849646}{0.02944216+}{0.62881825}{-0.12784369+}{0.76640509} \\
        & $u_{+, 2}$ = \Uloc{-0.14548208-}{0.62605359}{-0.33929403-}{0.68685619}{
            -0.74570464+}{0.17554615}{0.57544002-}{0.28631612} \\
        & $u_{-, 2}$ = \Uloc{-0.8076974+}{0.0317134}{0.09645005-}{0.58078959}{
            0.42100956+}{0.411546}{0.6703337-}{0.45170074} \\
        &  $v_{+, 2}$ = \Uloc{-0.20957887-}{0.87909618}{0.42114439-}{0.07683755}{
            -0.39442252-}{0.16642561}{-0.39541195+}{0.81263939} \\
        &  $v_{-, 2}$ = \Uloc{-0.01618801-}{0.74304507}{0.17034723-}{0.64699597}{
            -0.63284003+}{0.21710705}{0.7422453-}{0.03807824} \\
        \hline
        C 
        &  $u_{+, 1}$ = \Uloc{0.90604057-}{0.422824846}{0.00309542+}{0.01732210}{
            0.01759003-}{0.000477204}{0.23054939-}{0.97290150} \\
        & $u_{-, 1}$ = \Uloc{0.77125325+}{0.38295162}{0.09729126-}{0.49904999}{
            -0.47399433-}{0.18397243}{0.08759601-}{0.85662737} \\
        & $v_{+, 1}$ = \Uloc{-0.03365264+}{0.77810214}{-0.61909468-}{0.10072905}{
            -0.30046665-}{0.55058546}{-0.72284161+}{0.28995766} \\
        & $v_{-, 1}$ = \Uloc{0.30726796+}{0.91457831}{0.07859757+}{0.250909}{
            0.15206588-}{0.2144968}{-0.57405409+}{0.77545406} \\
        & $u_{+, 2}$ = \Uloc{0.02626481-}{0.66133538}{0.25064262+}{0.70648705}{
            -0.56068151+}{0.49757604}{-0.29640378+}{0.59177625} \\
        & $u_{-, 2}$ = \Uloc{0.42871255-}{0.60589777}{0.17130354-}{0.64788004}{
            -0.63831457-}{0.20407829}{0.74138121-}{0.03550296} \\
        & $v_{+, 2}$ = \Uloc{-0.72864047+}{0.07157325}{-0.13536488-}{0.66756025}{
            0.23703899+}{0.63857095}{0.65682528-}{0.32345048} \\
        & $v_{-, 2}$ = \Uloc{0.53434982+}{0.83729925}{-0.04025722+}{0.10853384}{
            -0.04081908-}{0.10832378}{-0.530004+}{0.84005686} \\
        \hline
        D 
        &  $u_{+, 1}$ =\Uloc{-0.26603916+}{0.20065589}{0.25447586-}{0.90785594}{
            -0.13280177-}{0.93344741}{-0.23732232-}{0.23391823} \\
        & $u_{-, 1}$ = \Uloc{-0.31678703+}{0.71576576}{-0.2903863+}{0.55045541}{
            -0.20823062-}{0.58648561}{0.21059044+}{0.7538742} \\
        & $v_{+, 1}$ = \Uloc{0.13185177+}{0.56539638}{-0.53320444+}{0.6153333}{
            -0.44095663-}{0.68447008}{-0.21038103+}{0.54110792} \\
        & $v_{-, 1}$ = \Uloc{-0.22733824-}{0.12247527}{0.77384188-}{0.57834755}{
            0.95104606+}{0.16978966}{0.14824766-}{0.21143675} \\
        & $u_{+, 2}$ = \Uloc{0.0780634-}{0.75379364}{-0.48091017-}{0.44093839}{
            -0.36269368-}{0.5423601}{0.75689899+}{0.03745222} \\
        & $u_{-, 2}$ = \Uloc{-0.55448735-}{0.59148658}{0.07652242-}{0.58037206}{
            -0.5333239-}{0.24135663}{0.20968476+}{0.78316339}  \\
        & $v_{+, 2}$ = \Uloc{0.44833607+}{0.08778421}{-0.87730783-}{0.14703629}{
            0.1319658-}{0.87970093}{0.05541789-}{0.4534756} \\
        & $v_{-, 2}$ = \Uloc{-0.23188006-}{0.18849329}{-0.78944085-}{0.53617634}{
            -0.92453077+}{0.23652648}{0.28207876-}{0.09863901} \\
        \hline
        E 
        &  $u_{+, 1}$ = \Uloc{-0.69009297-}{0.51913828}{0.25897705+}{0.43266387}{
            -0.27973694+}{0.4195406}{-0.71455207+}{0.48492082} \\ 
        & $u_{-, 1}$ = \Uloc{-0.69760973-}{0.58813487}{-0.12169375+}{0.39067719}{
            0.04608958+}{0.40658798}{-0.79572569+}{0.44652278} \\
        & $v_{+, 1}$ = \Uloc{-0.25224681+}{0.19387489}{0.41058323-}{0.85452062}{
            -0.92062667-}{0.2263418}{-0.31001755+}{0.07144959} \\
        & $v_{-, 1}$ = \Uloc{-0.5888309+}{0.80721348}{-0.0176661-}{0.03704691}{
            0.0128777+}{0.03897088}{0.98364188+}{0.17539693} \\
        & $u_{+, 2}$ =   \Uloc{0.62249645-}{0.54363084}{0.37530331-}{0.41965594}{
            -0.47107465-}{0.30830563}{0.74192191+}{0.36412662} \\
        & $u_{-, 2}$ = \Uloc{-0.52416883-}{0.60807282}{0.5702951-}{0.17394819}{
            -0.22456187-}{0.5523282}{-0.80246836-}{0.02345336} \\
        & $v_{+, 2}$ = \Uloc{0.06282152-}{0.88049157}{0.28027461-}{0.37713949}{
            -0.19875302-}{0.4257761}{-0.1154322+}{0.87514991} \\
        & $v_{-, 2}$ = \Uloc{-0.27681465-}{0.34651419}{0.79328171-}{0.4171399}{
            -0.8537797+}{0.27269357}{0.05274891+}{0.44035893} \\
        \hline
        F 
        & $u_{+, 1}$ = \Uloc{0.16746295+}{0.04160119}{-0.20111797+}{0.96424948}{
            0.922572-}{0.34508898}{-0.06652765-}{0.15921235}  \\ 
        & $u_{-, 1}$ = \Uloc{0.0898646-}{0.17188414}{-0.51690372+}{0.83378099}{
            -0.57135113-}{0.79745726}{-0.1011261-}{0.16550926} \\
        & $v_{+, 1}$ = \Uloc{0.78159121+}{0.00257971}{0.59336816-}{0.19241296}{
            0.04639178+}{0.62205814}{-0.29810347-}{0.72251353} \\
        & $v_{-, 1}$ = \Uloc{0.49866995-}{0.4247034}{0.56551407-}{0.50114782}{
            -0.65449515-}{0.37761277}{0.57368381+}{0.31611957} \\
        & $u_{+, 2}$ = \Uloc{-0.11310322-}{0.90210917}{-0.15532151-}{0.3863702}{
            0.07901707+}{0.40885573}{-0.39416074-}{0.81928664} \\
        & $u_{-, 2}$ = \Uloc{0.52536851-}{0.10764508}{-0.46311121+}{0.70564047}{
            0.84328645+}{0.03561493}{0.39396061-}{0.36386065} \\
        & $v_{+, 2}$ = \Uloc{0.44050915-}{0.09668913}{-0.71513996+}{0.53402035}{
            0.83571067-}{0.31335375}{0.31932476-}{0.31847888} \\
        & $v_{-, 2}$ = \Uloc{0.04940901+}{0.23028099}{-0.04022617+}{0.97103618}{
            0.41163049-}{0.88039182}{-0.14993899+}{0.18162843} \\
        \hline
    \end{tabular}
\end{table}

\section{Moments for two-site shift invariant operators \label{app:variance}}

Here we derive Eq.~\eqref{eq:var_omega_k} and its generalization to higher moments for
operators obeying two-site shift invariance, i.e., $\left[A, \T^2\right]=0$.
That is, we are aiming to express $m_{2p}(\omega)$ by the corresponding moments 
$m_{2p}^{(k)}(\omega)$ of the distribution of matrix elements of $\Proj_k A \Proj_k$
within the momentum $k$ subspace.
The basis of this subspace, in which the matrix elements $(\Proj_k A \Proj_k)_{mn}$ are
computed, is given by the eigenvectors $\ket{n}\in \Hil$ of $\U$ with $\Proj_k\ket{n}=\ket{n}$,
i.e. $k_n=k$.
More precisely, one has
\begin{align}
m_{2p}^{(k)}(\omega)
 = \frac{1}{N_k(\omega)}\sum_{m,n = 0}^{q^{2L}-1}|\left(\Proj_k A\Proj_k\right)_{mn}|^{2p}
 1_{I_{\Delta}(\omega)}(\omega_{mn}).
 \label{eq:higher_moment_k}
\end{align}
The normalization is given by
\begin{align}
N_k(\omega) = \frac{\Delta q^{4L}}{2\pi L^2} = \frac{N(\omega)}{L^2}.
\end{align}
as  the momentum $k$ subspace asymptotically has dimension $q^{2L}/L$
\cite{PinPro2007}.
Note that the sum~\eqref{eq:higher_moment_k} can
 be taken over all $m, n \in \{0, \ldots q^{2L-1}\}$ as the 
matrix element $(\Proj_k A \Proj_k)_{mn}$ is nonzero only if $k_m=k_n=k$.
Fixing arbitrary $m, n \in \{0,\ldots, q^{2L-1}\}$ this yields
\begin{align}
|A_{mn}|^{2p} & = \Big|\sum_{k=0}^{L-1} \bra{m}\Proj_k A \Proj_k \ket{n}\Big|^{2p} \\
& = \sum_{k=0}^{L-1} |(\Proj_k A \Proj_k )_{mn}|^{2p},
\label{eq:matrix_element_k}
\end{align}
where we used  $A = \sum_{k=0}^{L-1}\Proj_k A \Proj_k$.
Inserting Eq.~\eqref{eq:matrix_element_k} into Eq.~\eqref{eq:moments} finally gives
\begin{align}
m_{2p}(\omega)
& = \frac{1}{N(\omega)}\sum_{m,n = 0}^{q^{2L}-1}\sum_{k=0}^{L-1}|\left(\Proj_k A\Proj_k\right)_{mn}|^2
1_{I_{\Delta}(\omega)}(\omega_{mn}) \\
& = \frac{1}{L^2}\!\sum_{k=0}^{L-1}\frac{L^2}{N(\omega)}\!\sum_{m,n = 0}^{q^{2L}-1}\!|\left(\Proj_k A\Proj_k\right)_{mn}|^2
1_{I_{\Delta}(\omega)}(\omega_{mn}) \\
& =  \frac{1}{L^2}\sum_{k=0}^{L-1}m_{2p}^{(k)}(\omega),
\label{eq:var_omega_k_app}
\end{align}
which corresponds to Eq.~\eqref{eq:variance} for $p=1$ and allows for efficiently computing also higher
moments of the distribution of matrix elements.

\section{Asymptotical spectral function for eigenvectors of $\M_{\nu}$,
\label{app:fourier_transfo}}

In this section we briefly sketch the derivation of the asymptotical spectral function,
Eq.~\eqref{eq:f_analytic}, when $a$ in Eq.~\eqref{eq:extensive_sum} is an hermitian
eigenvector of the CPTP map $\M_{\nu}$ with real eigenvalue $\lambda$.
We write $\lambda=\exp(\ui \theta - \gamma)$ with $\theta \in \{0, \pi\}$ and $\gamma > 0$,
which yields $\tr\left(\M_{\nu}^t(a) a\right)/q = \exp(\left[\ui \theta - \gamma\right] t)$.
The Fourier transform, Eq.~\eqref{eq:spectral_function_asympt}, can then be evaluated as a
geometric series (omitting the dependence on $a$) as
\begin{align}
    |f_{\infty}(\omega)|^2 &  = 1 + \sum_{t=1}^{\infty}
     \ue^{\left[\ui (\theta + \omega) - \gamma\right] t} + 
     \ue^{\left[\ui (\theta - \omega) - \gamma\right] t} \\
    & = \frac{1}{1 - \ue^{\ui (\theta + \omega) - \gamma}} + \frac{1}{1 - \ue^{\ui (\theta - \omega) - \gamma}} - 1 \\
    & = \frac{1 - \ue^{2(\ui \theta - \gamma)}}{1 + \ue^{2(\ui \theta - \gamma)} - \ue^{\ui (\theta + \omega) - \gamma} - \ue^{\ui (\theta - \omega) - \gamma}} \\
    & = \frac{\ue^{-\ui \theta + \gamma} - \ue^{\ui \theta - \gamma}}
    {\ue^{-\ui \theta + \gamma} + \ue^{\ui \theta - \gamma} - \ue^{\ui\omega} - \ue^{-\ui \omega}} \\
    & = \frac{\sinh(\ui \theta - \gamma)}{\cos(\omega)-\cosh(\ui \theta - \gamma)},
\end{align}
from which Eq.~\eqref{eq:f_analytic} follows by distinguishing the cases $\theta = 0$,
corresponding to $\text{sign}(\lambda) = 1$, and  $\theta = \pi$,
corresponding to $\text{sign}(\lambda) = -1$, respectively.

\section{Parameters of local gates \label{app:parameters}}

In Table~\eqref{tab:local_unitaries} we report the matrices 
$u_{\pm},v_{\pm} \in \text{U}(2)$ entering 
Eq.~\eqref{eq:gates_parametrisation} for both half steps $i\in \{1, 2\}$
for the systems A-F.

\FloatBarrier


\end{document}